\documentstyle[aps,prl,epsfig,multicol]{revtex}
\begin{document}
\title{Structured Energy Distribution and Coherent AC Transport in Mesoscopic Wires}
\author{A. V. Shytov}
\address{Lyman Laboratory, Harvard University, Cambridge, MA 02138}
\address{L. D. Landau Institute for Theoretical Physics, 
    Kosygina str., 2, Moscow 117934, Russia}
\maketitle
\begin{abstract}
Electron energy distribution in a mesoscopic 
AC-driven diffusive wire
generally is not characterized 
by an effective temperature. 
At low temperatures, the distribution
has a form of a multi-step staircase, with
the step width equal to the field energy quantum. 
Analytic results for the field frequency high and low compared to
Thouless energy are presented, while the intermediate 
frequency regime is analyzed numerically. 
Manifestations in the tunneling spectroscopy and noise measurements 
are discussed.
\end{abstract}
\pacs{ PACS numbers:
05.30.-d,  
05.60.Gg 
}
\begin{multicols}{2}

In mesoscopic micron-size wires, at low temperatures, the electron-phonon 
energy relaxation time $\tau_{e-ph}$ 
can exceed the time $\tau_D$ of diffusion across 
the wire~\cite{Prober93}. 
In this regime, 
when electron cooling is primarily diffusive,
the system temperature has a very short thermal response
time to microwave radiation, determined mainly by the diffusion time $\tau_D$.
This makes such diffusion-cooled wires an excellent system for fast bolometric 
detection of terahertz radiation~\cite{Prober96,Karasik97}.

When the wire dimension, or electron temperature $T_e$, decreases further,
one arrives at the situation, demonstrated 
in Saclay experiments~\cite{Esteve-tunneling,Saclay_work}, when the electron-electron
energy relaxation is longer than the diffusion time $\tau_D$. 
When this happens, the electron dynamics can be considered as purely elastic.
Importantly, the effect of microwave field on electron energy
distribution in this case 
{\it is not described by an effective temperature}.
Since the energy of the field is absorbed in discrete quanta $\hbar\omega$,
and there is not enough time to redistribute it between electrons 
while they move around in the wire, one expects 
a staircase-like energy distribution to emerge, consisting of $\hbar\omega$ wide
steps. The steps will be pronounced when the temperature of electrons is 
below $\hbar\omega$. 

In this article, our goal is to develop a general approach that allows to analyze
such a situation. We shall focus on the regime when the field frequency is small
compared to the elastic scattering rate due to disorder, which corresponds to
the experimental situation~\cite{Prober96,Karasik97,Saclay_work}. Since in this case
the transport is diffusive on the field oscillation time scale $2\pi/\omega$,
the system can be described by momentum-averaged Greens functions,
similar to Usadel theory of disordered superconductors. 
We develop a general framework to analyze the energy distribution,
using Keldysh Greens functions, and then apply it to obtain analytic 
results in the two regimes, when the field frequency is small and large
compared to Thouless energy $E_T=\hbar/\tau_D$. In the intermediate regime,
$\omega \tau_D \sim 1$, we present numerical results.

The energy distribution of this form can be probed by 
tunneling spectroscopy, similar to the DC transport situation,
when a double step structure arises from 
mixing of the lead Fermi functions
at nonequal chemical potentials~\cite{Esteve-tunneling}.
Another manifestation we consider is in the shot noise, where
the structure of the energy distribution affects the noise power.

Before discussing the AC mesoscopic transport problem, 
let us recall, for comparison,
the basic facts about the DC transport. 
At low temperatures, when electron energy relaxation 
is slow, charge  transport is mainly controlled by elastic scattering due to
disorder. At high conductance, 
when the localization effects are negligible, 
the sample properties can be fully described by a scattering matrix.
All the statistics of transport in this case 
are controlled by a single parameter, the sample conductance
(see review~\cite{Beenakker97}). 

In the DC regime, the energy gained by an electron moving in a stationary 
electric field depends only on the total electron position displacement 
and is independent of the trajectory shape.
As a result, the electron energy distribution 
in a wire carrying a DC current does not depend on microscopic parameters
and is fully determined by the external voltage and
the wire geometry. The energy distribution in the DC case was explored 
by Nagaev~\cite{Nagaev92,Nagaev95}, 
who obtained a position-dependent mixture of 
two Fermi functions in the limit of slow energy relaxation,
and a single Fermi distribution with position-dependent effective
temperature in the limit of fast relaxation, and studied 
the effect on the shot noise.  
More recently, Nagaev~\cite{Nagaev02} has shown that 
the time-dependent current fluctuations of all orders can be expressed
through the nonequilibrium electron energy distribution.

In the AC case, in contrast, 
the energy gained by an electron depends on its whole trajectory. 
The AC transport can thus reveal additional information
about phase-sensitive effects 
in electron dynamics. 
It was pointed out by Lesovik and Levitov~\cite{Levitov-Lesovik} 
that even a slowly varying AC field leads 
to a new transport effect,  photon-assisted noise,
observed by Schoelkopf et al.~\cite{Schoelkopf} 
and recently by Glattli et al.~\cite{Glattli}.
The photon-assisted effects can be expected to become more drastic
and interesting when the field frequency is comparable to Thouless energy. 

In this paper,  a semiclassical treatment of ac transport in
mesoscopic wire is adopted. 
In a time-dependent external field, the uncertainty principle 
restricts  the ability to measure the energy of an electron as a function 
of time, and the time-dependent energy distribution function 
should be carefully defined. In this work, we employ the Wigner distribution 
function and show that in the presence of an AC field,
the time-averaged electron distribution consists
of $\hbar\omega$ steps. Subsequently, we discuss the 
manifestation of such distribution in tunneling 
spectroscopy~\cite{Esteve-tunneling} and in noise.

Let us now turn to the analysis of the energy distribution.
Non-equilibrium electrons in a diffusive conductor 
can be fully described by  the 
retarded and advanced Greens function $G^{R}(t,t', {\bf r}, {\bf r}')$ 
and $G^{A}(t, t', {\bf r}, {\bf r}')$, 
and Keldysh function $F(t,t', {\bf r}, {\bf r}')$ 
defined in~\cite{Keldysh}. In a diffusive conductor, many quantities 
can be calculated from momentum-averaged Greens function 
(i.e., taken at equal points in 
space, ${\bf r} = {\bf r}'$). Then, since for noninteracting electrons, 
the functions $G^{R}(t, t')$ and $G^{A}(t, t')$ are independent 
of the external fields, any quantity of interest can be expressed entirely in
terms of the Keldysh function $F(t, t', {\bf r})$.

The Keldysh function $F(t_1, t_2, {\bf r})$ contains information about both
energy distribution and the single-particle 
density of states. To separate the density
of states, we introduce the function $f(t_1, t_2, {\bf r})$ 
(see, {\it e.g.}, 
\cite{AltshulerReview}),
\begin{eqnarray}
F  = G_R  - G_A - 2  G_R  * f + 2 f * G_A \,, 
\end{eqnarray}
where $f * g$ stands for the convolution 
\begin{equation}
\label{convolution}
(f  * g)(t, t') 
= \int\limits_{-\infty}^{\infty} f (t, t'') g(t'', t') dt' \,.
\end{equation}
Then, the details of the electron spectrum 
are hidden in the Green's functions $G^{R,A} (t_1, t_2)$, 
while the function $f(t_1, t_2, {\bf r})$ 
depends only on the energy distribution. 

Consider a disordered wire of length $L$
with diffusion coefficient $D$, connected to
the leads which serve as a source of equilibrium electrons. The wire is
subject to electric field which we describe as an
external vector potential 
\begin{equation}
A_x(t) =  (cU / L\omega) \cos \omega t
\,,
\end{equation}
where $U$ is the amplitude of voltage across the wire, and $\omega$
is the frequency of the external field.
At frequencies low compared to the plasma frequency and Maxwell relaxation
rate, the longitudinal component of electric field is screened. 
Thus, the electric field can be described 
through the vector potential ${\bf A}({\bf r}, t)$ only.

The equal point Keldysh function $f(t_1, t_2, {\bf r})$ satisfies 
the two-time diffusion equation~\cite{Larkin,Altshuler,AltshulerReview}
\begin{equation}
\left\{ \frac{\partial}{\partial t_{+}}
+ D \left[ - i \nabla  
- \frac{e}{c} {\cal A}(t_1,t_2,{\bf r})  
   \right]^2
\right\} \, f(t_1, t_2, {\bf r}) = 0\,, 
\end{equation}
with $t_{+} = (t_1 + t_2)/2$ and 
${\cal A}(t_1, t_2,{\bf r}) = {\bf A}(t_1, {\bf r}) - {\bf A}(t_2, {\bf r})$.
This equation has to be supplemented with the boundary condition
at the leads:
\begin{equation}
f(t_1, t_2, x = 0) = f(t_1, t_2, x = L) = f_F (t_1-t_2)\,,
\end{equation}
where $f_F (t_1-t_2)$ is the Keldysh function of an equilibrium
Fermi gas.

It is convenient to rewrite the vector potential difference as
\begin{eqnarray}
& & {\cal A}_x(t_1, t_2, x) =
  A_x\left(t + \frac{\tau}{2}\right)
- A_x \left(t - \frac{\tau}{2}\right)
\\
\nonumber
& &= 2\, \frac{cU}{L\omega}\, \sin \omega t \sin \frac{\omega\tau}{2}
,\quad 
\tau=t_1-t_2 \,.
\end{eqnarray}
Passing to the Wigner representation, we 
replace $\tau$ by $i \partial_{\epsilon}$. 
Now, using the general operator relation
\begin{equation}
\exp\left(a \frac{\partial}{\partial \epsilon}\right) \Phi (\epsilon) =
\Phi (\epsilon + a)
\,,
\end{equation}
valid for an arbitrary function $\Phi(\epsilon)$, 
one arrives at
\begin{equation}
{\cal A}_x(t_1, t_2, x) = 
- \frac{ i c U}{L}\, \sin(\omega t) {\cal D}_{\omega} \,,
\end{equation}
where ${\cal D}_{\omega}$ is a finite difference operator
\begin{equation}
\label{fde-operator}
{\cal D}_{\omega} f(\epsilon)
= \frac{f (\epsilon + \omega / 2) - f (\epsilon - \omega / 2)}{\omega}\,.
\end{equation}
Next, we perform variable rescaling,
\begin{equation}
t\to t\omega,\quad
x\to x/L,\quad
\epsilon \to\epsilon/eU
\end{equation}
i.e., measure time in the units of $\omega^{-1}$,
$x$ in the units of $L$, and energy in the units of $eU$.
After that, the equation for the distribution function becomes
\begin{equation}
\label{eq-diff-dimensionless}
\left[
\frac{\partial }{\partial t} - \frac{1}{\omega \tau_D}
\left( \frac{\partial}{\partial x} - \sin t {\cal D}_{\omega}\right)^2
\right]\, f (t, \epsilon, x) = 0 \,, 
\end{equation}
where $\tau_D = L^2 / D$ is the time of diffusion across the wire. 

For a wire connected to the leads,
the boundary condition to the Eq.~(\ref{eq-diff-dimensionless})
takes the form
\begin{equation}
\label{boundary-condition}
f(t, \epsilon, x=0) = f(t, \epsilon, x=1) = n_F (\epsilon)\,,
\end{equation}
where $n_F (\epsilon)$ is a Fermi distribution.
For a wire that has been in equilibrium with the leads before the
field was turned on, the initial condition is
\begin{equation}
\label{initial-condition}
f(t=0, \epsilon, x) = n_F (\epsilon) 
\end{equation}
for any $x$.

Note that the operator ${\cal D}_{\omega}$ in Eq.~(\ref{eq-diff-dimensionless})
relates the values of the distribution function $f (\epsilon)$
only for the energies that differ by $\pm \frac12 \hbar \omega $.
Thus, at zero temperature
the singularity at $\epsilon = 0$ in the boundary
condition~(\ref{boundary-condition}) propagates
to the energies $\epsilon_n = \frac12 n \hbar \omega$.
Also, since the Fermi distribution
$n_F (\epsilon)$ is flat for $\epsilon < 0$ and $\epsilon > 0$,
the distribution function
is also flat for $\epsilon_n < \epsilon < \epsilon_{n + 1}$.
Thus, the profile of $f (t, \epsilon, x)$ 
is a series of steps at energies $n \hbar \omega / 2$.
In fact, 
only the even steps 
($\epsilon_{2n} = n \hbar \omega$) corresponding to
absorption of individual field quanta, survive after averaging over time.

For slow field, $\omega \tau_D \ll 1$, one may
neglect the time derivative in Eq.~(\ref{eq-diff-dimensionless}).
The solution to Eq.~(\ref{eq-diff-dimensionless})
is then
\begin{eqnarray}
\label{slow-general}
f(t, \epsilon, x) &=& \left[
(1 - x) \exp \left(x {\cal D}_\omega \sin t\right) 
\right.
\\
\nonumber
&+& 
\left.
x \exp\left((x - 1){\cal D}_\omega \sin t \right)
\right] n_F (\epsilon)
\,. 
\end{eqnarray}
The exponent of the operator ${\cal D}_\omega$
can be found using Fourier transform in the energy domain. 
The result is
\begin{equation}
\label{kernel}
\exp (\alpha {\cal D}_\omega) \Phi (\epsilon)
= \sum\limits_{n=-\infty}^{\infty} \Phi (\epsilon - n \omega / 2)
J_n (2\alpha / \omega)\,, 
\end{equation}
where $\Phi(\epsilon)$ is an arbitrary function, 
$J_n(x)$ is the Bessel function of $n$th order, 
and the sum runs over all integer $n$. 
As we discuss below, one is mostly interested 
in the time-averaged distribution function, because
it can be directly measured by tunneling spectroscopy. 
Substituting Eq.~(\ref{kernel}) into Eq.~(\ref{slow-general}),
and using the formula 
$\overline{J_{2k}(2a \sin t)} = J_k^2 (a)$
to average over $t$, one finds, 
with the original units restored,
\begin{equation}
\label{f-zero}
\bar{f}_0(\epsilon, x) = \left(1 - \frac{x}{L}\right) 
                          F_0\left(\epsilon, x\right)
                       + \frac{x}{L}\, F_0 \left(\epsilon, {L-x}\right)\,,
\end{equation}
where
\begin{equation}
F_0(\epsilon, x) = 
\sum\limits_{k > \epsilon / \omega}^{\infty} J_k^2 
                     \left( \tilde u \right)
,\quad
\tilde u=\frac{x eU}{L\hbar\omega}
\,.
\end{equation}
(We mention that an expression of the form 
similar to (\ref{slow-general}) for nonaveraged
function $f$ was used by Altshuler {\it et al.} 
in the calculation of time-dependent 
noise~\cite{Altshuler94}.)

To interpret the result (\ref{f-zero}), note that the work performed by a slow field
on electrons that travel a distance $x$ along the wire, is  $eUx/L$. 
Then, according to~\cite{Tien-Gordon}, the probability of 
absorbing $n$ field quanta is $J_n^2(\tilde u)$.
The factor $(1-x/L)$ in Eq.~(\ref{f-zero}) is the number of the
electrons at cross-section $x$ 
coming from the left lead (cf.~\cite{Nagaev92}). The second 
term in Eq.~(\ref{f-zero}) describes electrons that come from the right 
lead. 

In the limit $\omega \ll eU$, one may use the asymptotic form of 
the Bessel function at $x/\omega \sim n \gg 1$, which gives
\begin{equation}
\label{f-very-slow}
F_0 (\epsilon, x) = \left\{
\begin{array}{lc}
1  , & {\tilde \epsilon < -1} \\
{\frac{1}{\pi}\,\cos^{-1}\tilde \epsilon} 
             , &  {|\tilde \epsilon| < 1} \\
0 , & {\tilde \epsilon > 1}
\end{array}
\right.
,\quad \tilde \epsilon = \frac{L\epsilon}{xeU}
\,, 
\end{equation}
This result can also be derived by time averaging the two-step distribution
found in~\cite{Nagaev92} over the time-dependent voltage difference.

For fast field, $\omega \tau_D \gg 1$, 
the first term in Eq.~(\ref{eq-diff-dimensionless})
is the most important. Thus, the electron distribution
is almost time-independent. (The time-dependent part of 
$f(t, \epsilon, x)$ is proportional to $(\omega \tau_D)^{-1}$.)
Projecting out the time-dependent part of the distribution function, 
i.e., averaging the Eq.~(\ref{eq-diff-dimensionless})
over field period,  one finds the equation for the time-averaged
electron distribution:
\begin{equation}
\label{laplace-discrete}
\left[\frac{\partial^2}{\partial x^2} +
\frac{1}{2} {\cal D}_{\omega}^2\right] \bar{f}(\epsilon, x) = 0\,.
\end{equation}
Using Fourier transform with respect to energy, one 
can find the solution of Eq.~(\ref{laplace-discrete}) 
in a closed integral form.
It is more instructive, however, to solve
this equation in the limit $\hbar \omega \ll eU$, where one 
can replace ${\cal D}_{\omega}$ by $\partial_\epsilon$. 
This brings Eq.~(\ref{laplace-discrete}) to the form of a Laplace's equation
in the two-dimensional strip $0 < x < 1$, $-\infty < \epsilon < \infty$. 
Using the function
\begin{equation}
w = \exp(\pi i (x + i \sqrt{2} \epsilon))
\,,
\end{equation} 
one can  conformally 
map this strip onto half-plane $\mathop{\rm Im}\nolimits w > 0$. 
The boundary condition~(\ref{boundary-condition}) on the line 
$\mathop{\rm Im}\nolimits w = 0$ is, for $T_e = 0$:
\begin{equation}
\label{bc-laplace}
\bar{f}_\infty(w) = 
\left\{
   \begin{array}{lc}
   0, &  |\mathop{\rm Re}\nolimits w| < 1 \, , \\ 
   1, &   |\mathop{\rm Re}\nolimits w| > 1 \, 
   \end{array}
\right.
\,.
\end{equation}
This boundary value problem is solved by
an imaginary part of the analytic function
\begin{equation}
\bar{f}_\infty(w) = {\rm Im}\, \frac{1}{\pi} \, \log\left(\frac{1 - w}{1 + w}\right)
\end{equation}
Restoring the original dimensional units, one obtains 
\begin{equation}
\label{f-infinity}
\bar{f}_{\infty}
(\epsilon, x) 
= \frac1{\pi}\,{\rm cotan}^{-1}\left(\frac{\sinh (\pi\sqrt{2}\epsilon/eU)}{\sin (\pi x/L)}\right)
\end{equation}
Unlike the low frequency solution (\ref{f-zero}) 
with $F_0(\epsilon, x)$ given 
by Eq.~(\ref{f-very-slow}), 
the high frequency distribution Eq.~(\ref{f-infinity}) is non-zero 
at large $\epsilon$. Physically, this difference is due to tha fact
that maximal energy 
that can be gained from the field is given by the bias 
amplitude $eU$ 
only in the DC regime. 
In contrast, in a varying field, there is a finite probability that electron 
diffuses several times back and forth in phase with the field, 
thus gaining energy that exceeds $eU$.

\begin{figure}[t]
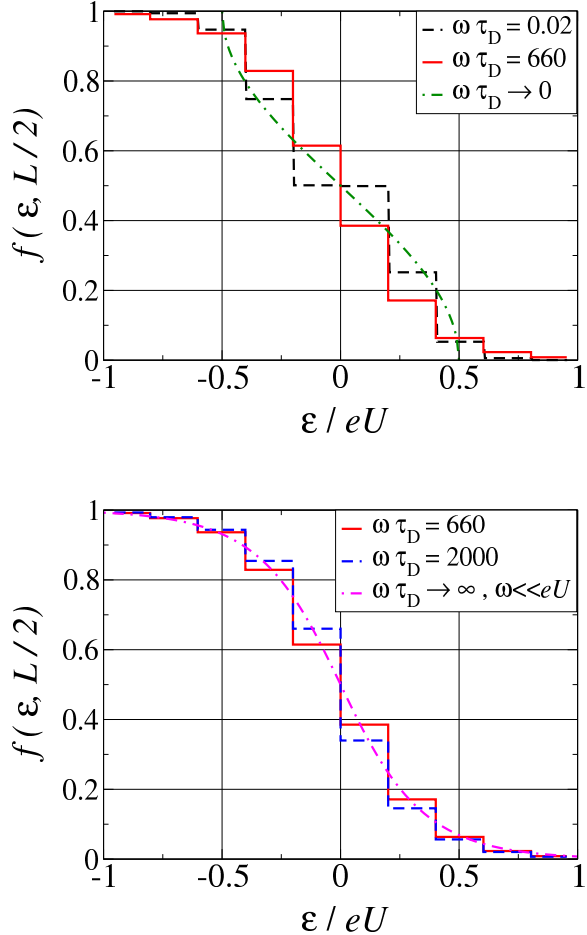

\centerline{
\begin{minipage}[t]{3in}
\vspace{1pt}
\centering
\includegraphics[width=3in]{fig2a_wire.eps}
\end{minipage}
\hspace{-0.1in}
}
\centerline{
\begin{minipage}[t]{3in}
\vspace{0.2in}
\centering
\includegraphics[width=3in]{fig2b_wire.eps}
\end{minipage}
\hspace{-0.1in}
}
\vspace{0.15cm}
\caption{
\label{fig2}
The energy distribution at the wire midpoint $x = L/2$, calculated
numerically for $\hbar\omega=0.2 eU$, for different values
of $\omega\tau_D$. 
The height of the $\hbar\omega$ steps is a function of external voltage
and other system parameters. The envelope of the staircase 
structure changes as the frequency of the field increases, demonstrating
a crossover from the slow to the fast field regime.}
\end{figure}

Eqs.~(\ref{f-zero}),~(\ref{f-very-slow}) and~(\ref{f-infinity})
are correct in the limit $\omega \ll eU$.
For finite values of $\omega / eU$ these results approximate 
the average profile of the step-like
electron distribution. To obtain the distribution at intermediate
frequencies, $\omega\tau_D\simeq 1$,
Eq.~(\ref{eq-diff-dimensionless}) was solved numerically for different
values of $eU / \omega$. The results are shown on Fig.~\ref{fig2}.
From that figure one may see that a crossover from 
the low-frequency behavior to
the high-frequency behavior occurs at $\omega \tau_D \sim 100$.
To understand the origin of this large numerical factor, 
note that the electron distribution relaxes at $t\to\infty$, 
according to the diffusion equation,  as $\exp(-\mu t)$, where $\mu$
is the lowest non-zero eigenvalue of a diffusion operator: 
$\mu = \pi^2 / \tau_D$. Assuming that the crossover occurs when
the relaxation time is of order of field period, $2\pi / \omega$, 
one finds $\omega \tau_D \sim 2 \pi^3 \approx 60$, 
in a qualitative agreement with the numerical data.

To characterize the electron distribution, one may compute its second
moment, or shot noise. 
For low frequencies, the answer can be derived
just by averaging over time the well known result 
for the shot noise~\cite{Nagaev92}:
\begin{equation}
\label{shot-noise-zero}
S_0 = \frac{2}{3} e G \overline{|U(t)|}
= \frac{4}{3\pi}\, e GU \approx 0.424 e GU\,,
\end{equation}
where $G$ is a conductance of the wire.
For high frequencies, one may neglect the time dependence of the distribution
function and compute the noise as~\cite{Nagaev92}
\begin{equation}
S_\infty
= 4 e G U \int\limits_{-\infty}^{\infty}\int\limits_{0}^{1}
d\epsilon\, dx\,
\bar{f}_{\infty}(\epsilon, x)  (1 - \bar{f}_{\infty}(\epsilon, x))\,.
\end{equation}
Evaluating the integral, one arrives at
\begin{equation}
\label{shot-noise-inf}
S_\infty = \frac{21 \sqrt{2}}{2 \pi^3} \,\zeta(3)\, eGU
\approx 0.575\, eGU\,,
\end{equation}
where $\zeta(3)$ is Riemann zeta-function. \
The value given by Eq.~(\ref{shot-noise-inf}) is higher 
than that in Eq.~(\ref{shot-noise-zero}), 
because of the contributions of high energy tails
of the distribution~(\ref{f-infinity}).

The electron distribution can also be probed by superimposing 
a DC voltage $U_{\rm dc}$, which splits every step into 
two, separated by $eU_{\rm dc}$. When this splitting becomes 
equal to $\hbar \omega$, the ``resonance'' between two steps
leads to a singularity in the shot noise and higher current moments, 
predicted for short wires in~\cite{Levitov-Lesovik}
and observed in~\cite{Schoelkopf}. 

\begin{figure}
\epsfxsize=0.95\hsize
\epsfbox{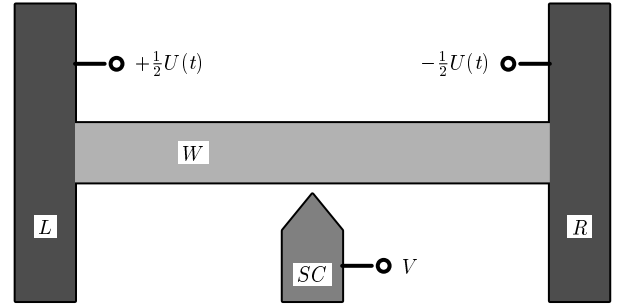}
\caption{
\label{fig1}
Schematic representation of the measurement~\protect\cite{Esteve-tunneling},
in which the energy distribution in a mesoscopic wire (W) is probed by 
a NS tunneling junction between the wire and the superconducting
probe (SC). By changing voltage $V$ across the junction, 
one may scan the energy of tunneling electrons. 
AC voltage $U(t)$ applied between the leads ($L$ and $R$),
creates a  nonequilibrium energy distribution in the wire.
}
\end{figure}

Finally, we discuss how the structure in the energy distribution considered above 
will manifest itself in the tunneling spectroscopy 
measurement~\cite{Esteve-tunneling,Saclay_work}.
In this setup, a short segment of the sample at point ${\bf r}_0$ is
connected to a metallic probe by a tunneling junction 
(see Fig.~\ref{fig1}). The tunneling density of states of the probe 
must contain a sharp feature that serves as a ``pointer'',  
allowing to probe electron distribution. 
(In~\cite{Esteve-tunneling}, the probe was a superconductor, 
and the BCS square root singularity near the superconducting gap was used.)
Applying the probing voltage $V$ to the junction, 
allows one to scan the energy domain.
The  electron energy distribution
in the sample is then extracted 
from the $I-V$ curve of the tunnel
junction. 

To derive the relation between the tunneling 
current and the energy distribution in a nonequilibrium system, 
one may use the standard tunneling Hamiltonian
\begin{equation}
\hat{H}_{T} (t) = \sum_{{\bf p}, {\bf k}} T_{{\bf p}, {\bf k}} 
    a_{{\bf p}}^+(t) \, b_{\bf k}(t) \, + {\rm h. c.}
\end{equation}
(see, e.g., \cite{Mahan}). Here $a_{\bf p}(t)$ and $b_{\bf k}(t)$
are electron operators of the sample and the probe. 
The tunneling current operator in this formalism is
\begin{equation}
\hat{I}_T(t) = i \sum_{{\bf p}, {\bf k}} T_{{\bf p}, {\bf k}}
    a_{{\bf p}}^+(t) \, b_{\bf k}(t) \, + {\rm h. c.}\,.
\end{equation}
The average tunneling current $I(t)$ can be found from Kubo formula:
\begin{equation}
I(t) = i \int\limits_{-\infty}^{t} dt' 
\left\langle \left[\hat{H}_{T}(t'), \hat{I}_T(t) \right]\right\rangle \,. 
\end{equation}
Using the Keldysh approach~\cite{Keldysh}, 
one may express the tunneling current $I(t)$ in terms of
retarded and advanced Green's function $G^{R(A)}(t_1, t_2)$
and Keldysh function $F(t_1, t_2)$ defined in~\cite{Keldysh}:
\begin{eqnarray}
\label{I-tunn-F}
I(t) \propto 
\mathop{\rm Re}\nolimits \left.\left[ F_s * G_p^R - F_p * G_s^R \right]
\right|_{t' = t}
\end{eqnarray}
(subscripts $s$ and $p$ denote the sample and the probe,
all functions are taken at space point  ${\bf r}_0$, where the tunneling
occurs). The star denotes the convolution defined by Eq.~(\ref{convolution}). 

It is now convenient to express all quantities in 
frequency domain. 
In the noninteracting system, 
the functions $G^{R,A} (t_1, t_2)$ depend only on the difference 
of their arguments $t_1 - t_2$. 
Their Fourier components are related through the Kramers-Kronig relation
to the density of states $\nu(\omega)$ defined by
\begin{equation}
\pi \nu (\omega) = 
\mathop{\rm Im}\nolimits G^{R} (\omega) =
- \mathop{\rm Im}\nolimits G^{A} (\omega) \,. 
\end{equation}
The function $f(t_1, t_2)$ in general depends on both arguments.
We define its Fourier transform as 
\begin{equation}
f_\omega(\epsilon) = \int dt\, dt' e^{i \epsilon (t - t')}\, 
e^{i\omega (t + t') / 2}\, f(t, t')\,. 
\end{equation}
Thus, $f_\omega(\epsilon)$ is a Fourier transform of the Wigner distribution
\begin{equation}
f(\epsilon, t) = 
\int f (t + \tau / 2, t - \tau / 2) e^{i\epsilon \tau} \, d\tau
\,.
\end{equation}
Then, the Fourier components of the tunneling current~(\ref{I-tunn-F})
are given by
\begin{eqnarray}
I_\omega 
\propto\!
\int\limits_{-\infty}^{\infty}\!\! {d\epsilon}\, 
\nu_s\left(\epsilon + \frac{\omega}{2}\right) 
\nu_p\left(\epsilon + \frac{\omega}{2}\right) 
\bigl[f_{s, \omega} (\epsilon) - f_{p,\omega} (\epsilon)\bigr]\, .
\end{eqnarray}
In an equilibrium system, the function $f_\omega(\epsilon)$
coincides with the Fermi distribution~\cite{Keldysh}: 
$f_\omega (\epsilon) = 2\pi n_F (\epsilon) \delta(\omega)$.
Thus, $f(\epsilon, t)$ can be identified 
as the conventional energy distribution function.

The probing voltage  $V$ simply shifts
energy levels of the probe by $eV$. Thus, the tunneling current is: 
\begin{eqnarray}
\label{I-tunn-f}
I_\omega (eV)  &\propto& 
\int {d\epsilon}\,
\nu_s \left(\epsilon + \frac{\omega}{2}\right) 
\nu_p \left(\epsilon + \frac{\omega}{2} + eV\right) \nonumber
\\
& & \times 
\left[ f_{s, \omega} (\epsilon) - f_{p, \omega} (\epsilon + eV) \right]
\,. 
\end{eqnarray}

Eq.~(\ref{I-tunn-f}) shows that in general, the relation between the
tunneling current and energy distribution function is not
straightforward, because all factors in~(\ref{I-tunn-f})
depend on the current frequency $\omega$.
(One cannot extract the energy distribution from tunneling
measurements if both $\nu_s (\epsilon)$
and $\nu_p (\epsilon)$ are flat.) The $\omega$-dependence of the
product $\nu_s (\epsilon + \omega / 2) \nu_p (\epsilon + \omega / 2 + eV)$
leads to  a non-local relation between
energy distribution and tunneling current. This non-locality 
does not allow to measure Wigner function directly. 
However, for $\omega = 0$, the average tunneling current is proportional to
the average electron distribution:
\begin{equation}
\label{I-tunn-0}
\bar{I} (eV)  \propto  
\int d\epsilon
\nu_s (\epsilon) \nu_p (\epsilon + eV)
\left[\bar{f_s} (\epsilon) - \bar{f_p} (\epsilon + eV)\right] \,,  
\end{equation}
where the bar denotes averaging over one field period.  
Thus, the tunneling spectroscopy technique of \cite{Esteve-tunneling}
allows to measure directly the time-averaged electron energy distribution.

The effect of finite temperature can be understood as follows. 
At finite electron temperature $T_e$, the singularity in the
boundary condition~(\ref{boundary-condition}) is smeared.
The resulting smearing of the step-like structure 
is negligible for $\hbar \omega > T_e$.
For a wire length $L = 1\mu m$, the electron mean free path
$l = 10 nm$, and Fermi velocity $v_F = 10^{7}$ m/s
the diffusion time  is $\tau_D = L^2/v_Fl = 10^{-11}$ s.
The fast field regime occurs for $\omega \tau_D > 100$,
{\it i.e.}, for the frequencies $f = \omega / 2\pi > 1$ THz.
The optimal voltage amplitude is of the order of $10 \hbar \omega$,
{\it i.e.} $V \sim 50\,{\rm mV}$. 
The temperature must be less than $\hbar\omega / k_B$,
{\it i.e.}, $T < 10\,{\rm K}$. These conditions appear
realistic, and can be fulfilled in an experiment.

The simple theory presented in this paper does not take into account
the effect of non-equilibrium density fluctuations
in the wire on  the density of states of the superconducting 
probe through long-range Coulomb forces~\cite{Narozhny}.
The treatment of this 
effect for an AC field   
is beyond the scope of this paper.
However, since the non-equilibrium fluctuations are 
also due to periodic field,  the resulting tunneling current 
will also have singularities at quantization energies 
$n \hbar \omega$, although the shape of these singularities 
may differ from simple steps. 

To summarize, the energy distribution of a mesoscopic 
AC-driven wire is not described by an effective temperature. 
The most interesting feature of the electron 
distribution is the $\hbar\omega$
steps due to quantization of
absorbed energy. Those steps are present at 
the external field frequencies larger than 
electron temperature. The calculated envelope 
in the fast-field regime distribution
is qualitatively different from that of the slow-field regime. 
Experimental manifestations of these effects experimentally 
are discussed.

I am grateful to D.~Esteve, H.~Pothier,
M.\,V.~Feigel'man, L.\,S.~Levitov and E.\,G.~Mishchenko 
for stimulating discussions.
This research was supported by the Russian Ministry of Science under the
program ``Physics of quantum computing,'' by
RFBR grant 98-02-19252 and NSF grant PHY99-07949.

\end{multicols}
\end{document}